\documentclass[global]{svjour}
\usepackage{amssymb}
\usepackage{amsfonts}
\usepackage{amsmath}

\input{tcilatex}
\begin{document}

\title{Exact solutions to a massive charged scalar field equation in the
magnetically charged stringy black hole geometry and Hawking radiation}
\author{ I. Sakalli\inst{1} \and A. Al-Badawi\inst{2}}
\institute{Department of Physics, Eastern Mediterranean University, Gazimagosa, North
Cyprus, Mersin 10, Turkey%
\email{izzet.sakalli@emu.edu.tr}%
\and Department of Physics, AL-Hussein Bin Talal University, Ma'an, Jordan 
\email{dr.ahmadbadawi@ahu.edu.jo}%
.}
\dedication{}
\offprints{}
\mail{}
\maketitle

\begin{abstract}
Exact solutions of a massive complex scalar field equation in the geometry
of a Garfinkle-Horowitz-Strominger (stringy) black hole with magnetic charge
is explored. The separated radial and angular parts of the wave equation are
solved exactly in the non-extreme case. The angular part is shown to be an
ordinary spin-weighted spheroidal harmonics with a spin-weight depending on
the magnetic charge. The radial part is achieved to reduce a confluent Heun
equation with a multiplier. Finally, based on the solutions, it is shown
that Hawking temperature of the magnetically charged stringy black hole has
the same value as that of the Schwarzschild black hole.
\end{abstract}

\keywords{Klein-Gordon Equation, Charged String Black Hole, Confluent Heun
Equation, Hawking Radiation.}

\section{INTRODUCTION}

Exact solution to a sourceless charged massive scalar field equation in the
Kerr-Newman black hole geometry was first found by Wu and Cai \cite{R1}.
Later on, they have extended their previous study \cite{R1} to exact
solution of a scalar wave equation in a Kerr-Sen black hole geometry \cite%
{R2}, which reduces to the electrically charged
Garfinkle-Horowitz-Strominger (GHS) geometry in the limit of vanishing
rotation parameter. However, no one has considered to solve the problem of
the massive complex scalar field equation in the magnetically charged GHS
geometry, yet. So we think that it might be useful to fill this absence in
the literature. The main goal of this paper is to show that exact solutions
to a massive charged scalar field equation on magnetically charged GHS
background is somewhat different than the electrically charged GHS geometry 
\cite{R2} and investigate Hawking evaporation of scalar particles.

Recall that according to the laws of black hole mechanics \cite{R3}, Hawking 
\cite{R4,R5} proved that a stationary black hole can emit particles from its
event horizon with a constant temperature proportional to the surface
gravity.\ Since Hawking's this pioneering study, many methods have been
proposed in order to calculate the Hawking radiation during the last three
decades, see for instance \cite{R6,R7,R8,R9,R10}. One of the commonly used
methods is the method of Damour-Ruffini-Sannan (DRS) \cite{R11,R12}. This
method is applicable to any Hawking's problems in which asymptotic behaviors
of the wave equation are known near the event horizon.

According to the DRS method, it is plausible to investigate the exact
solutions of the scalar wave equation in order to get more accurate
calculations about Hawking radiation. In general, obtaining an exact
solution to the wave equation in a given geometry is so difficult. On the
other hand, the main reason of a researcher's motivation is indeed such
difficulties. Today, a wider perspective of the properties and physics of
black holes can be acquired by studying \ other types of black hole
solutions appearing in string theory. Of particular interest is considering
GHS \cite{R13} black hole, which is a member of a family of solutions to
low-energy limit of string theory. It is discovered when the field content
of Einstein-Maxwell theory is enlarged to include a dilaton field $\phi $,
which couples to the metric and the gauge field, non-trivially. This causes
the charged stringy black holes to differ significantly from the
Reissner-Nordstr\"{o}m (RN) black hole. Very recently, the Hawking radiation
of the GHS black hole (in the string frame) has been studied by using the
method of cancellation of anomalies at the horizon \cite{R14}.

In this paper, we shall not discuss the extreme case since it does not have
the characteristics of a black hole solution. In generally, the crucial
equation, which plays an essential role in the calculation of the Hawking
radiation is the radial equation. In the non-extreme case, it is shown that
the radial equation reduces to a confluent Heun equation. Although the Heun
differential equations are less known than the hypergeometric family in the
literature, due to the necessities of their using in various physical
problems, they have been intensively attracting much interest. One may refer
to \cite{R15,R16} in order to see the applications of the Heun equations to
many modern physical problems. Heun equations also appear in the quantum
mechanical problems of general relativity. For instance, it can be seen that
more recently Al-Badawi and Sakalli \cite{R17} have shown that the angular
part of the Dirac equation in the rotating Bertotti-Robinson geometry is
solved in terms of the confluent Heun functions.

The paper is organized as follows: In Sec. II, a brief overview of the GHS \
black hole solution is given. Next, we separate a massive magnetically
charged scalar field equation on the GHS spacetime into the angular and
radial parts. The solutions to radial equation in non-extreme case is
devoted to Sec. III. Next, we shall employ the DRS method to discuss Hawking
radiation in Sec. IV. Finally, we draw our conclusions.

\section{GHS SPACETIME AND SEPARATION OF KLEIN-GORDON EQUATION ON IT}

In the low-energy limit of string field theory, the four-dimensional action
(in Einstein frame) describing the dilaton field $\phi $ coupled to a $U(1)$
gauge field is

\begin{equation}
S=\int d^{4}x\sqrt{-g}(R-2(\nabla\phi)^{2}-e^{-2\phi}F^{2})
\end{equation}

where $F_{\mu \upsilon }$ is the Maxwell field associated with a $U(1)$
subgroup of $E_{8}\times E_{8}$ or Spin(32)/$Z_{2}$. In the presence of a
magnetic charge the dilaton cannot be constant and the static, spherically
symmetric solutions designated with GHS black holes \cite{R13} are given by

\begin{equation}
ds^{2}=-(1-\frac{2M}{r})dt^{2}+\frac{dr^{2}}{1-\frac{2M}{r}}+r(r-\frac {%
Q^{2}e^{-2\phi_{0}}}{M})(d\theta^{2}+\sin^{2}\theta d\varphi^{2}),
\end{equation}

with

\begin{equation*}
e^{-2\phi}=e^{-2\phi_{0}}(1-\frac{Q^{2}e^{-2\phi_{0}}}{Mr}),
\end{equation*}

\begin{equation}
F_{\theta\varphi}=Q\sin\theta,
\end{equation}

\bigskip where $\phi _{0}$\ is the asymptotic constant value of the dilaton
and $Q$ is the magnetic charge. For electric charge case one can generate
the solutions by applying the duality transformations

\begin{equation}
\widetilde{F}_{\mu \upsilon }\rightarrow \frac{1}{2}e^{-2\phi }\epsilon
_{\mu \upsilon }^{\rho \sigma }F_{\rho \sigma }\text{ \ \ \ and \ \ \ }\phi
\rightarrow -\phi .
\end{equation}

Note that this transformation does not modify the geometry (2). In this
case, the solution for the dilaton and electromagnetic field are given by

\begin{equation*}
e^{2\phi}=e^{-2\phi_{0}}(1-\frac{Q^{2}e^{-2\phi_{0}}}{Mr}),
\end{equation*}

\begin{equation}
F_{rt}=\frac{Q}{r^{2}},
\end{equation}

where $Q\ $refers now to the electric charge. Although the Einstein metric
(2) is the same for both electrically and magnetically charged black holes,
one of the black hole solutions should be considered while separating the
Klein-Gordon equation in this geometry. This is because the electromagnetic
four-vector potentials are different for the both black holes. Here, we are
interseted in metric (2) together with fields (3), and for simplicity, we
set $\phi _{0}=0$. It is easy to derive the electromagnetic four-vector
potential of the magnetically charged stringy black holes as follows

\begin{equation}
A_{\mu}=-Q\cos\theta\delta_{\varphi}^{\mu},
\end{equation}

Let us compare solution (2) to RN which represents the solution of a charged
black hole given by the following metric 
\begin{equation}
ds^{2}=-(1-\frac{2M}{r}+\frac{Q}{r^{2}})dt^{2}+\frac{dr^{2}}{1-\frac{2M}{r}+%
\frac{Q}{r^{2}}}+r^{2}(d\theta ^{2}+\sin ^{2}\theta d\varphi ^{2}),
\end{equation}

One can see some important differences immediately. First of all, contrast
to the pure Einstein-Maxwell theory there is only one horizon at $r=2M$\ .
In fact the $R^{2}$ part of metric (2) is identical to the Schwarzschild
black hole. This implies that also the surface gravity \cite{R18} coincides
with Schwarzschild

\begin{equation}
{\LARGE \kappa=}\left[ \underset{r\rightarrow2M}{\lim}(-\frac{1}{4}%
g^{tt}g^{ij}g_{tt,i}g_{tt,j})\right] ^{\frac{1}{2}}=\frac{1}{4M}.
\end{equation}

Important differences appear in the angular part. There is a curvature
singularity (spacelike), hidden inside the horizon, when the radius of
two-sphere vanishes at $r=\frac{Q^{2}}{M}$. Since the DRS method concerns
the asymptotic behaviors of the scalar waves near the event horizon, this
spacelike singularity does not make any trouble on using of this method.
Another difference with respect to the RN solution concerns the extremal
configuration. Here it is given by $\left\vert Q\right\vert =\sqrt{2}M$
instead of the condition $\left\vert Q\right\vert =M$\ for the RN black
holes. In the extreme limit, the area of the event horizon $A=8\pi M(2M-%
\frac{Q^{2}}{M})$ shrinks to zero and turns out to be singular. Namely, a
naked singularity appears, so the solution is no longer a black hole.

In curved spacetime, a massive charged test scalar field $\Phi $ with mass $%
\mu $ and charge $q$ obeys the covariant Klein-Gordon equation \cite{R1}.
The massive scalar wave function $\Phi $ in metric (2) can be separated as $%
\Phi (r,t,\theta ,\varphi )=R(r)G(\theta )e^{i(m\varphi -\omega t)}$, in
which the angular part $G(\theta )$ satisfies the following equation

\begin{equation}
G^{\prime \prime }+\cot \theta G^{\prime }+\left[ \frac{\lambda }{M}-\frac{%
(m+qQ\cos \theta )^{2}}{\sin ^{2}\theta }\right] G=0.
\end{equation}

where $\lambda $ is a separation constant. (Throughout the paper, a prime
denotes the derivative with respect to its argument.)

Letting $\pounds =\frac{\lambda }{M}=l(l+1)-p^{2}$, one can see that $%
G(\theta )e^{im\varphi }$ is nothing but a spin-weighted spherical harmonics 
$_{p}Y_{lm}(\theta ,\varphi )$. Here, it should be highlighted that the
spin-weight appears as $p=qQ$ \cite{R19}, which is seen as a differentness
when we compare it with the electrically charged case \cite{R2}. Basically,
this result is the consequence of the four-vector potential (6) involving an
angular term. One should notice that by the modified separation constant $%
\pounds ,$ the magnetic charge is carried into the radial equation, which is
going to be discussed in the next section.

\section{REDUCTION OF THE RADIAL TO A CONFLUENT HEUN EQUATION}

In this section, we shall show that the radial part of the massive complex
scalar field equation is in fact a confluent Heun equation. Using metric (2)
as being a background for the covariant Klein-Gordon equation, we see that
the separated radial part of the massive charged test scalar field $\Phi $
with mass $\mu $ is governed by the following equation

\begin{equation*}
(r-2M)(Q^{2}-rM)R^{\prime \prime }+\left[ Q^{2}-rM-M(r-2M)\right] R^{\prime
}+\left[ \lambda -\mu ^{2}r(Q^{2}-rM)+\right.
\end{equation*}

\begin{equation}
\left. \frac{r^{2}\omega ^{2}(Q^{2}-rM)}{r-2M}\right] R=0,
\end{equation}

Making the following coordinate transformation 
\begin{equation}
r=2M-Dx,
\end{equation}

where $D=\frac{2M^{2}-Q^{2}}{M}$, letting $k=\sqrt{\omega ^{2}-\mu ^{2}}$
(assuming that $\omega >\mu $ ) and substituting 
\begin{equation}
R(r)=x^{2i\omega x}e^{-ikDx}H(x),
\end{equation}

into differential equation (10), then we can reduce it to a confluent form
of Heun equation \cite{R20}

\begin{equation*}
H^{\prime\prime}+(\alpha+\frac{\beta+1}{x}+\frac{\gamma+1}{x-1})H^{\prime}+
\end{equation*}

\begin{equation}
\frac{1}{x(x-1)}\left\{ \eta +\frac{\beta }{2}+\frac{1}{2}(\beta +1)(\gamma
-\alpha )+\left[ \frac{\alpha }{2}(\beta +\gamma +2)+\delta \right]
x\right\} H=0,
\end{equation}

which shows that it has two singular points at $x=0,1$ and one irregular
singular point at the infinity, $x=\infty $ \cite{R15,R16}. It yields the
following specific parameters:

\begin{align}
\gamma & =0,\text{ \ \ \ \ }\alpha =-2ikD,\text{ \ \ \ \ }\beta =4i\omega M,%
\text{ \ \ \ }  \notag \\
\text{\ }\delta & =-2DM(\omega ^{2}+k^{2})\text{ \ \ \ \ }\eta =-(\pounds %
+\delta ),
\end{align}

In the literature, see for instance \cite{R21}, there are special
transformations, which make possible to express the Heun functions in terms
of ordinary special functions. Today, all well-known transformations from
confluent Heun functions to other special functions are listed in the famous
computer package, \emph{MAPLE} 10 and its higher versions. Adapting Maple's
notation for the confluent Heun functions, we obtain the following canonical
solutions of equation (13) 
\begin{equation}
H(x)=C_{1}HeunC(\alpha ,\beta ,\gamma ,\delta ,\eta ;x)+C_{2}x^{-\beta
}HeunC(\alpha ,-\beta ,\gamma ,\delta ,\eta ;x).
\end{equation}

The convergent Taylor series expansion of the confluent Heun functions with
respect to the independent variable $x$ around regular singular point $x=0$
(i.e. around event horizon $r=2M$) is obtained using the known three-terms
recurrence relation \cite{R15,R16} and initial conditions:

\begin{equation}
HeunC(\alpha ,\beta ,\gamma ,\delta ,\eta ;0)=1,
\end{equation}

and

\begin{equation}
\left. HeunC^{\prime }(\alpha ,\beta ,\gamma ,\delta ,\eta ;x)\right\vert
_{x=0}=\frac{(1+\beta )(\gamma -\alpha )+\beta +2\eta }{2(1+\beta )}.
\end{equation}

\section{HAWKING RADIATION OF SCALAR PARTICLES}

It is easy to see from equations (11) and (15) that the radial solutions
(12) near to the horizon behave asymptotically as

\begin{equation}
R(r)\sim C_{1}(r-2M)^{2i\omega M}+C_{2}(r-2M)^{-2i\omega M},
\end{equation}

Therefore, just outside the horizon ($r>2M$) two linearly independent
solutions exist:

1- The outgoing wave solution

\begin{equation}
\Phi ^{out}\rightarrow C_{1}(r-2M)^{2i\omega M}e^{-i\omega t}\text{ }%
_{qQ}Y_{lm}(\theta ,\varphi ).
\end{equation}

2- The ingoing wave solution

\begin{equation}
\Phi ^{in}\rightarrow C_{2}(r-2M)^{-2i\omega M}e^{-i\omega t}\text{ }%
_{qQ}Y_{lm}(\theta ,\varphi ),
\end{equation}

in which $\omega $ is assumed to be a positive. As $r\rightarrow \infty $
the outgoing wave has an infinite number of oscillations and therefore
cannot be straightforwardly extended to the interior region of the black
hole in contrast with the ingoing wave. On the other hand, the outgoing wave
can be analytically extended from outside into the interior of the black
hole by the lower half complex $r$-plane

\begin{equation}
\left( r-2M\right) \rightarrow (2M-r)e^{-i\pi }.
\end{equation}

According to the DRS method \cite{R11,R12}, a correct wave describing a
particle flying off of the black hole is

\begin{equation}
\Phi =N_{\omega }\left[ \Theta (r-2M)\Phi _{r>2M}^{out}+e^{4\pi M\omega
}\Theta (2M-r)\Phi _{r<2M}^{out}\right] ,
\end{equation}

where $\Theta $\ is the conventional Heaviside function and $N_{\omega }$ is
a normalization factor. Since $\Phi ^{out}$ differs from $\Phi ^{in}$ as a
factor $(r-2M)^{-4i\omega M}$, the above complexified analytical treatment
(21) requires to put a difference factor of $e^{4\pi M\omega }$ into
equation (22). Thus we can derive the relative scattering probability of the
scalar wave at the event horizon

\begin{equation}
\mathcal{R}=\left\vert \frac{\Phi _{r>2M}^{out}}{\Phi _{r<2M}^{out}}%
\right\vert ^{2}=e^{-8\pi M\omega },
\end{equation}

and the resulting radiation spectrum of scalar particles are obtained as
follows

\begin{equation}
\left\vert N_{\omega }\right\vert ^{2}=\frac{\mathcal{R}}{1-\mathcal{R}}%
=(e^{8\pi M\omega }-1)^{-1}.
\end{equation}

Using the formal definition of the radiation spectrum \cite{R18}, one can
easily read the Hawking temperature as%
\begin{equation}
T_{H}=\frac{1}{8\pi M}=\frac{\kappa }{2\pi }.
\end{equation}

This result shows that the statistical Hawking temperatures of the
Schwarzschild black hole and the GHS black hole (independent of its charge
type) are the same.

\section{CONCLUSION}

In this paper, our target was to investigate exact solutions of a massive
complex scalar field equation in the magnetically charged string black hole
background, which is referred to as the GHS black hole. After getting the
exact solution, we have applied the method of DRS to derive the Hawking
radiation of the magnetically charged GHS black holes. On the other hand, we
should state that the present calculation of the Hawking temperature
reproduces the result expected from more general analyses, which has been
recently made in \cite{R22} due to the inspection of the form of the metric
in Eq.(2).

The separated angular part is obtained in terms of the spin-weighted
spheroidal harmonics with a spin-weight, which peculiarly depends on the
product of the charges, $qQ$. On the other hand, the separated radial part
is successfully reduced to a confluent form of the Heun equation. After
using the initial conditions of the confluent Heun functions, which are
obtained by the virtue of the Taylor series expansion around the event
horizon, the asymptotic behaviors of the ingoing and outgoing scalar waves
are defined near to the horizon. Here, after using the DRS method, we have
shown that the thermal property of the magnetically charged stringy black
holes closely resemble to the electrically charged stringy black holes.
Namely, a charged stringy black hole shares similar quantum thermal effect
as the Schwarzschild spacetime exhibits. Nevertheless, there might be a way
to reveal differences between these black hole radiations. To the end this,
one may consider the classical approximation \cite{R9} to compute the
Hawking Radiation. This is an equivalent procedure to calculating the
Bogoliubov coefficients relating two vacua: The vacuum for a quantum field
near the horizon is not same to the observer's vacuum at infinity. Briefly,
it is a procedure to compute the reflection and absorption coefficients of a
wave by the black hole. However, the coefficient for reflection by the black
hole can be best calculated whether one may find a relevant transformation
between the confluent Heun functions $HeunC(\alpha ,\beta ,\gamma ,\delta
,\eta ;x)$\ and $HeunC(\alpha ,\beta ,\gamma ,\delta ,\eta ;1/x)$ such as in
the hypergeometric functions \cite{R9}. But, a transformation from $x$ to $%
1/x$ of the argument of the confluent Heun functions does not exist in the
literature. The main difficulty in making such a transformation arises due
to the fact that the point $x=0$ is a regular singular point in contrast to
the point $x=\infty $, which is an irregular singular point. In summary,
nowadays useful asymptotic expansions of the confluent Heun functions are
still open questions.

Finally, it should be emphasized that to properly study Hawking radiation in
a selected geometry, the backreaction of the quantum effects must be taken
into account. However, such an attempt resorts to the complete theory of
quantum.
\begin{verbatim}
ACKNOWLEDGEMENTS
\end{verbatim}

We would like to thank the anonymous referee for drawing our attention to an
incorrect statement in the paper, and for her/his valuable comments and
suggestions. Special thanks to Dr. E.C.Terrab and Dr. P.Fiziev for their
helpful comments on the confluent Heun functions.

\end{document}